\documentclass[prl,twocolumn,superscriptaddress,showpacs]{revtex4}
\usepackage{amsmath,amssymb,graphicx}

\begin{document}
\title{

%Dynamics
Kinetics of stimulated polariton scattering in planar
microcavities: Evidence for a dynamically self-organized optical
parametric oscillator }

\author{A. A. Demenev}
 \affiliation{Institute of Solid State Physics RAS, Chernogolovka, 142432 Russia}

\author{A. A. Shchekin}
 \affiliation{Institute of Solid State Physics RAS, Chernogolovka, 142432 Russia}

\author{A. V. Larionov}
 \affiliation{Institute of Solid State Physics RAS, Chernogolovka, 142432 Russia}

\author{S. S. Gavrilov}
 \affiliation{Institute of Solid State Physics RAS, Chernogolovka, 142432 Russia}
 \affiliation{A. M. Prokhorov General Physics Institute RAS, Moscow, 119991
Russia}

\author{V. D. Kulakovskii}
 \affiliation{Institute of Solid State Physics RAS, Chernogolovka, 142432 Russia}

\author{N. A. Gippius}
 \affiliation{A. M. Prokhorov General Physics Institute RAS, Moscow, 119991 Russia}
   \affiliation{LASMEA, UMR 6602 CNRS, Universit\'{e} Blaise Pascal,
   %24 av.~des Landais, 63177
Aubi\`ere, France}

\author{S. G. Tikhodeev}
\affiliation{A. M. Prokhorov General Physics Institute RAS, Moscow, 119991 Russia}
   \affiliation{LASMEA, UMR 6602 CNRS, Universit\'{e} Blaise Pascal,
   %24 av.~des Landais, 63177
Aubi\`ere, France}

\date{November 26, 2007}
%\date{August 21, 2007}
%\date{June 16, 2007}

\begin{abstract}
We demonstrate for the first time the strong temporal hysteresis
effects in the kinetics of the pumped and scattered polariton
populations in a planar semiconductor microcavity under a
nano-second-long pulsed resonant (by frequency and angle)
excitation above the lower polariton branch. The hysteresis
effects are explained in the model of multi-mode scattering when
the bistability of the nonlinear pumped polariton is accompanied
by the explosive growth of the scattered polaritons population.
Subsequent self-organization process in the nonlinear polariton
system results in a new --- dynamically self-organized --- type of
optical parametric oscillator. \\
\end{abstract}


\maketitle

%%%%%%%%%%%%%%%%%%%%%%%%%%%%%%%%%%%%%%%%%%%%%%%%%%%%%%%%%%%%%%%%

%\textit{Introduction.}--

A giant stimulated polariton-polariton scattering is one of the
most striking features in the optical response of planar
microcavities (MCs). The scattering was firstly observed in
GaAs-based MCs with InGaAs quantum wells (QWs) in the active layer
under a cw excitation at wave-vector $\mathbf{k}_\mathrm{p}$ close
to the inflection point of lower polariton (LP) branch
$\omega_\mathrm{LP}(\mathbf{k})$, when the scattering exhibits an
unusually low (smaller than 400~W/cm$^2$)
threshold~\cite{Stevenson2000,Tartakovskii2000,Baumberg2000}.
Specifically, such excitation results in the strong parametric
scattering into states positioned approximately on
$\omega_\mathrm{LP}(\mathbf{k})$ with $\mathbf{k}_\mathrm{s} = 0$
and $\mathbf{k}_\mathrm{i} = 2\mathbf{k}_\mathrm{p}$, called
signal and idler, respectively. The effect was theoretically
described in terms of four-wave mixing or parametric
scattering~\cite{Ciuti2001,Whittaker2001,Savvidis2001}.
Subsequent studies~\cite{Kulakovskii2001,Butte2003,Gippius2004}
have shown that the shift of the excitation from the inflection
point of the LP dispersion is not followed by the corresponding
shift of the stimulated scattering along the LP branch
characteristic for the four wave mixing. Instead, the scattering
goes on to the same states with $\mathbf{k} \sim 0$ and $\sim
2\mathbf{k}_\mathrm{p}$. The energy conservation is then fulfilled
by the shift of the signal and the idler much above the LP branch.

The stability analysis of the single macro-occupied pump mode as
well as numerical simulations of the polariton scattering
indicate~\cite{Kulakovskii2003,Gippius2004,Gippius2004a,Gippius2005a}
that such unusual behavior can result from the interplay between
two instabilities in the resonantly excited MC: bistability of the
pumped polariton mode intensity with respect to the external pump,
and its parametric instability with respect to the decay into
multiple scattered polaritons in a {\it wide} range of
$\mathbf{k}$.

The single-mode optical bistability in MC for
pumping at $\mathbf{k}_\mathrm{p}=0$  has been observed and explained
theoretically within one-mode optical parametric oscillator (OPO)
model~\cite{Baas2004}. The bistability of the scattered signal at
pumping near inflection point was also found and explained within
three-mode OPO model~\cite{Baas2004a}.

Different possible regimes of the above-threshold OPO have been
analyzed theoretically within the three-mode
approximation~\cite{Whittaker2005,Wouters2007}.  However
three-mode OPO model cannot determine the signal and idler wave
vectors $\mathbf{k}_{s,i}$ which are selected by the parametric
process above threshold. In the system with a specific
dispersion (containing
an inflection point), the interplay between the
pump single mode bistability and its multi-mode parametric
instability can result in a regime where essentially
multi-mode coupling between the ensemble of lower polaritons
plays the decisive role in the formation of the signal and idler
\cite{Gippius2004,Gippius2004a,Gavrilov2007}. We will refer to
this model as a \textit{dynamically self-organized (DSO)} OPO.

Dramatic change of LP scattering pattern from the figure-of-eight
shape corresponding to spontaneous regime at lower pump
intensity~\cite{Ciuti2001,Langbein2004} to that directed along
$\mathbf{k} \sim 0 $ and $\sim 2 \mathbf{k}_\mathrm{p}$ for pump
powers above the parametric scattering threshold observed
recently~\cite{Krizhanovskii2007} supports the DSO OPO model.
However, the direct evidence of multi-mode self-organized nature
of the scattering demands the study of the LP system dynamics. The
previous time-resolved studies of the LP dynamics under ps-long
excitation pulses~\cite{Langbein2004} have revealed a well
pronounced figure-of-eight distribution of the final LP scattering
states in low excitation regime, but only a small narrowing of the
LP momentum distribution with time at higher excitation.

In this Letter we report the optical studies of kinetics of the LP
system under ns-long pump pulses and discover for the first time
the strong hysteresis effects in the kinetics of the optical
response of the pumped as well as of the scattered MC polaritons,
thus directly demonstrating the behavior, predicted within the DSO
OPO model~\cite{Gippius2004,Gippius2004a,Gavrilov2007}.

%\textit{Experimental.}--
The MC structure has been grown by a metal organic vapor phase
epitaxy. The Bragg reflectors are composed of 17 (20) repeats of
$\lambda/4$ Al$_{0.13}$Ga$_{0.87}$As/AlAs layers in the top
(bottom) mirrors. The $3/2 \lambda$ GaAs cavity contains six 10-nm
thick In$_{0.06}$Ga$_{0.94}$As/GaAs quantum wells (QWs). The Rabi
splitting is $\Omega \sim  6$~meV. A gradual variation of an
active layer thickness along the sample provides a change in the
photon mode energy $E_C$ and, accordingly, in the detuning
$\Delta$  between the exciton $\hbar \omega_X(\mathbf{k}=0)$ and
photon  $\hbar \omega_C(\mathbf{k}=0)$ mode energies. Experiments
have been carried out on several spots of the same sample with
$\Delta$ in the range from -1.5 to -2~meV.

The sample was placed into an optical cryostat with controlled
temperature. A pulsed Ti-sapphire laser with pulse duration of
$\sim 1$~ns, line full width at half maximum (FWHM) of $\sim
0.7$~meV, and pulse repetition of 5~kHz has been used for the
excitation of the MC at the angle $14^\circ$ relative to the
cavity normal. The pump beam has been focused onto a spot with
diameter of ~100~$\mu$m. The kinetics of angular distribution of
PL signal $I(\mathbf{k},t)$ from the MC has been detected in a
wide solid angle around the cavity normal by the streak camera
with spectral, angular, and time resolution of 0.28~meV,
1$^\circ$, and 70~ps, respectively. The transmission signal
$I_\mathrm{tr}(\mathbf{k}_\mathrm{p},t)$ has been detected with
the same streak camera.

\begin{figure}
\begin{center}
\includegraphics[width=0.6\linewidth]{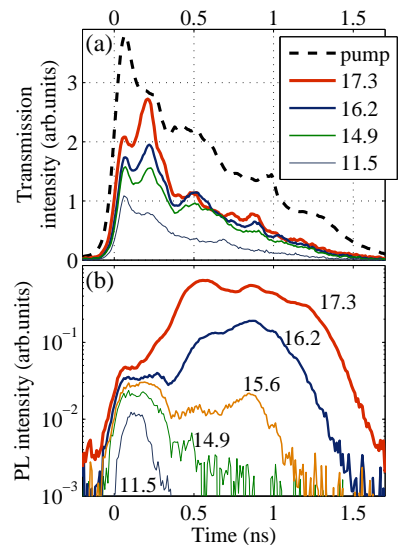}
\caption{ \label{fig1}
(Color online) (a) Time dependences of the pump pulse
$I_{P}(\mathbf{k}_\mathrm{p},t)$ and of the transmission
$I_\mathrm{tr}(\mathbf{k}=\mathbf{k}_\mathrm{p},t)$ at different
excitation densities. (b) Kinetics of the LP emission intensity
$I(\mathbf{k}=0,t)$ at different excitation densities. Numbers
show the peak pump intensity $P$ in kW/cm$^2$.
}
\end{center}
\end{figure}

%\textit{Experimental results.}--
The pump pulse profile $I_{P}(t)$ is shown in Fig.~\ref{fig1}a
(dashed curve). The magnitude of $I_P(t)$ determines the intensity
of the external electric field outside the  MC
$|E_\mathrm{ext}(t,\mathbf{k}_\mathrm{p})|^2\propto I_P(t)$. The
intensity builds up during the first 100~ps and then decreases
monotonically (by about 3 times to $t=1$~ns). The pulses with a
circular ($\sigma^+$) polarization and a spectral FWHM of $\sim$
0.7 meV excite the MC with $\Delta=-2$ meV about 0.5 meV above the
LP dispersion branch at $\mathbf{k}_\mathrm{p} = (k_{px}, k_{py})
= (1.96,0)$~$\mu$m$^{-1}$. Figure~\ref{fig1}b shows the recorded
kinetics of MC emission normal to its plane, $I(\mathbf{k}=0,t)$.
The spectra are recorded in the $\sigma^+$ polarization. The
emission in $\sigma^-$ is about two orders of magnitude smaller.

The MC emission $I(\mathbf{k} = 0,t)$ is proportional to
$|E_\mathrm{QW}(\mathbf{k}=0,t)|^2$, the intensity of the
$\mathbf{k} = 0$ harmonic of the electric field on QW
\textit{inside} MC. Figure~\ref{fig1}b clearly shows that
$I(0,t)$ and thus the time dependence of $|E_\mathrm{QW}(0,t)|^2$
differs significantly from the exciting pulse shape. At low $P =
11.5$ kW/cm$^2$, the signal reaches its maximum only slightly
later (by $\sim $ 50 ps) than the pumping pulse and then decreases
quickly, by one order of magnitude at $t \sim 0.35$~ns when the
pump intensity is still about 60\% of the maximum. At
$P>15.5$~kW/cm$^2$, the signal behavior changes drastically. After
its marked decrease  (together with the pump) in the range of $
t=0.2-0.35$~ns the signal starts to grow and reaches the second
maximum at $t \sim 0.85$~ns already at the excitation pulse fall
off. The intensity of this maximum grows by more than two orders
of magnitude in the range of $P$ between 14.9 and 17.2~kW/cm$^2$,
i.e., threshold-like.

The recorded time dependences of the MC transmission at the pump
angle $I_\mathrm{tr}(\mathbf{k}_\mathrm{p},t)$ are shown in
Fig.~\ref{fig1}a (solid curves).
$I_\mathrm{tr}(\mathbf{k}_\mathrm{p},t)$ is proportional to the
intensity of the $\mathbf{k} =\mathbf{k}_\mathrm{p}$ harmonic of
the QW electric field $|E_\mathrm{QW}(\mathbf{k}_\mathrm{p}),t|^2$
inside MC. Again we see clearly that
$I_\mathrm{tr}(\mathbf{k}_\mathrm{p},t)$ and thus
$|E_\mathrm{QW}(\mathbf{k}_\mathrm{p})|^2$ do not directly follow
the exciting field
$|E_\mathrm{ext}(\mathbf{k}_\mathrm{p},t)|^2\propto I_P(t)$. At $P
\lesssim 11.5$ kW/cm$^2$  $I_\mathrm{tr}$ and, thus,
$|E_\mathrm{QW}(\mathbf{k}_\mathrm{p})|^2$ is a monotonous
superlinear function of
$|E_\mathrm{ext}(\mathbf{k}_\mathrm{p})|^2$ on both the up and
down going parts of the excitation pulse, the maxima of
$I_\mathrm{tr}(t)$ and $I_P(t)$ and, hence, those of
$|E_\mathrm{QW}(\mathbf{k}_\mathrm{p}),t|^2$ and
$|E_\mathrm{ext}(\mathbf{k}_\mathrm{p},t)|^2$ nearly coincide with
each other. However, the monotonous dependence
$I_\mathrm{tr}(I_P)$ becomes distorted with increasing $P$.
Figure~\ref{fig1}a shows that $I_\mathrm{tr}$ starts to
demonstrate a narrow second peak in the range of nearly constant
exciting field $|E_\mathrm{ext}(\mathbf{k}_\mathrm{p})|^2$ at
$t\sim 0.2$~ns: The growth of
$|E_\mathrm{QW}(\mathbf{k}_\mathrm{p})|^2$ starts at
$t=0.12\pm0.02$~ns, continues about 0.1~ns, and gives way to its
sharp decrease at $t=0.23\pm0.02 $~ns. The duration of the
increase and decrease in
$|E_\mathrm{QW}(\mathbf{k}_\mathrm{p})|^2$ is close to an
available time resolution of our detecting system of 70~ps. The
second peak of $|E_\mathrm{QW}(\mathbf{k}_\mathrm{p})|^2$ grows
quickly with $P$ and shifts slightly towards the pulse onset.

\begin{figure}
\begin{center}
\includegraphics[width=0.9\linewidth]{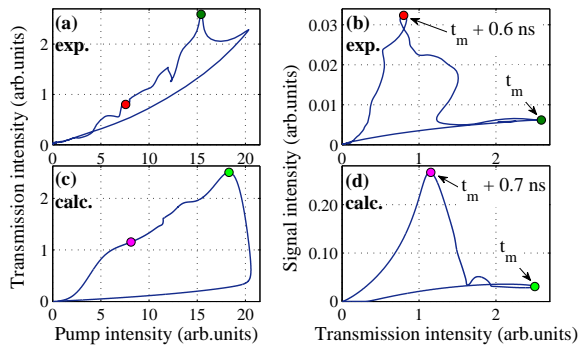}
\caption{ \label{fig2}
(Color online) Measured (top panel) and  calculated  (bottom
panel) dependences of transmission intensity on pump intensity and
PL intensity on transmission intensity at $P = 16.2$ kW/cm$^2$.
Circles mark the characteristic points with maxima of transmission
(at $t = t_m$) and $\mathbf{k}=0 $ emission (approximately 650~ns
later).
}
\end{center}
\end{figure}
Time dependences   $I_P(t)$, $I(\mathbf{k}=0,t)$, and
$I_\mathrm{tr}(\mathbf{k}_\mathrm{p}=0,t)$ in Fig.~\ref{fig1} can
be redrawn as implicit functions $I_\mathrm{tr}(I_P)$ and
$I_S(I_\mathrm{tr})$ presenting the dependences of the inner field
$|E_\mathrm{QW}({\mathbf{k}=\mathbf{k}_\mathrm{p}})|^2$ on the
exciting field
$|E_\mathrm{ext}({\mathbf{k}=\mathbf{k}_\mathrm{p}})|^2$ and of
the $\mathbf{k} = 0$ harmonic of the  QW electric field on
$|E_\mathrm{QW}({\mathbf{k}=\mathbf{k}_\mathrm{p}})|^2$,
respectively, for each $P$. Figures~\ref{fig2}a and b show the
resulting dependences at  $P = 16.2$ kW/cm$^2 > P_\mathrm{thr}$.
Both the experimentally measured dependences of the
$\mathbf{k}=\mathbf{k}_\mathrm{p}$ electric field inside MC on the
external field
$E_\mathrm{QW}|_{\mathbf{k}=
\mathbf{k}_\mathrm{p}}(E_\mathrm{ext}|_{\mathbf{k}=\mathbf{k}_\mathrm{p}})$
and of the $\mathbf{k}=0$ electric QW field inside MC on that at
$\mathbf{k}=\mathbf{k}_\mathrm{p}$
$E_\mathrm{QW}|_{\mathbf{k}=0}(E_\mathrm{QW}|_{\mathbf{k}_\mathrm{p}})$
acquire jumps and hysteresis behaviour.

These experimental results  find their qualitative explanation in
the framework of the system of a semi-classical Gross-Pitaevskii
type equation for QW excitonic polarization
$\mathcal{P}(\mathbf{k},t)$ and a Maxwell equation for
$E_\mathrm{QW}(\mathbf{k},t)$ in response to the driving external
field far from the
MC~\cite{Gippius2004,Gippius2004a,Gavrilov2007}. According to this
theoretical model, the dynamics of the stimulated parametric
scattering in the planar MCs has a following scenario. Its start
is initiated by a single-mode instability of the pumped mode at
$\mathbf{k}=\mathbf{k}_\mathrm{p}$, which results in the jump of
$|E_\mathrm{QW}(\mathbf{k}_\mathrm{p})|^2$ and transfers this mode
into the region of its strong instability with respect to the
parametric LP-LP scattering at once into a large range of
$\mathbf{k}$. That provides an explosive growth of LP population
in a wide $\mathbf{k}$-space region, mainly around $\mathbf{k}=0$,
on one hand, and causes the abrupt decrease in the driven mode
population, on the other hand. The formation of an OPO with a
three dominating macrooccupied modes with signal and idler at
$\mathbf{k}=0$ and $2\mathbf{k}_\mathrm{p}$ occurs due to a
dynamical self-organization in the multi mode scattering and takes
a long -- hundreds of ps -- time.

The calculated dynamics of the MC optical response within the
model of Refs.~\cite{Gippius2004,Gippius2004a,Gavrilov2007} at the
excitation slightly above the hard excitation threshold are
displayed in Fig.~\ref{fig2}c and d. The system of
Gross-Pitaevskii and Maxwell equations has been numerically solved
for experimental-like time dependence of $E_\mathrm{ext}(t)$. As
seen from Fig.~\ref{fig2}c,d, the model demonstrates the
hysteresis in dependences of $I_\mathrm{tr}$ on $I_P$ (panel c)
and of $I(\mathbf{k} = 0)$ on $I_\mathrm{tr}$ (panel d) similar to
the experimental ones. The considered model  takes into account
only coherent scattering processes (nondiagonal components of the
LP density matrix) and neglects important scattering processes
like LP--phonon or LP--free carriers. Nevertheless, it
demonstrates the hysteresis in a qualitative agreement with the
experiment.

The hysteresis of the dependence of $I_\mathrm{tr}$ on
$I_\mathrm{p}$ results from the blue shift of LP eigenenergy. The
increase of an overall LP population shifts the driven mode
frequency towards the pump frequency, and results in increased
transmissivity. Even when the pump intensity has been fallen down,
the system retains considerable signal population that keeps it
close to the resonance. At the same time, the hysteresis of
$E_\mathrm{QW}(\mathbf{k}_\mathrm{s}\sim0)$ vs.
$I_\mathrm{tr}\propto |E_\mathrm{QW}(\mathbf{k}_\mathrm{p})|^2$
reveals a more complicated nature of the studied system. Indeed,
the model does not presume any characteristic time rather than the
lifetime of cavity polaritons (${\sim} 3$ ps), hence the long time
of the signal developing ($\tau \sim 10^2$ ps) might come from
only the collective phenomena caused by numerous inter-mode
scattering processes. Since the evolution to the ``three-mode''
($\mathbf{k} = \{ 0, \, \mathbf{k}_\mathrm{p}, \,
2\mathbf{k}_\mathrm{p} \}$) state involves a lot of modes, that
state appears as an essentially collective formation. Moreover,
the \textit{eventual} state differs from stable``three-mode'' OPO
solution even in the case of stationary pump, which may be proved
by performing the stability analysis similar to that discussed
in~\cite{Whittaker2005}. The actual stability of the three-mode
pattern is maintained by the presence of numerous weak
``above-condensate'' modes, so the whole system occurs to be
highly correlated, i.e., it demonstrates a new -- dynamically
self-organized -- type of OPO.

\begin{figure}
\begin{center}
\includegraphics[width=0.9\linewidth]{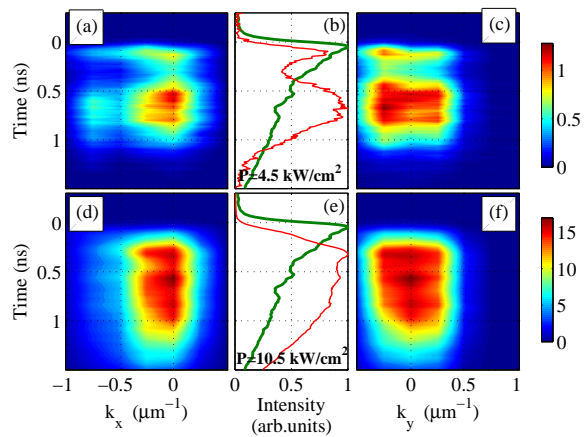}
\caption{ \label{fig3}
(Color online) Time dependences of a $k$-distribution of LP
emission at $P= 4.5$ (upper panels) and 10.5 (lower panels)
kW/cm$^2$. Left (a,d) and right (c,f) panels display the $k_x$
dependence at $k_y=0$ and $k_y$-dependence at $k_x=0$,
respectively. The thick and thin solid lines in the central panels
(b,e) show, respectively,  the pulse and MC emission profiles.
}
\end{center}
\end{figure}
To support this scenario of DSO OPO experimentally, the LP
scattering dynamics in a wide range of $\mathbf{k}$ has been
measured by recording time dependences of an angle distribution of
MC emission. Panels (a,d) in Fig.~\ref{fig3} display
$k_x$-distribution at $k_y=0$ whereas panels (c,f) display $k_y$
distribution at $k_x$=0 for the point on the MC sample with a
smaller $P_\mathrm{thr}\sim 3.75$ kW/cm$^2$ for two excitation
densities $P= 4.5 $ and 10.5~kW/cm$^2$, i.e., slightly and well
above $P_\mathrm{thr}$. $I_P(t)$ and $I(\mathbf{k}=0,t)$ are given
in panels (b,e). The $k-$distribution of the emission is symmetric
in $k_y$ direction but shows a well pronounced asymmetry in the
direction of exciting pulse ($k_x$). The maximum emission in the
very beginning of the pulse is at $k_x\sim -0.4~\mu$m$^{-1}$ and
then during $t\sim 0.1$~ns it shifts to $\mathbf{k}=0$. This
behavior is well expected. The phonon-assisted scattering
dominating at low LP densities cannot provide the LP relaxation to
$\mathbf{k}=0$ because of comparable magnitudes of LP life time
and phonon assisted scattering time~\cite{Bloch1997,Tassone1997}.
The effective LP relaxation at higher densities appears due to the
onset of LP-LP scatterings.

Figures~\ref{fig3}a and c show that dynamics at $P= 4.5$ kW/cm$^2$
($\sim 20$ \% above $P_{thr}$). The scattering developing after
the bistable transition at $t\sim 0.4$~ns occurs into a wide range
of wavevectors $k_x$ between $-0.7$ and $+0.3~\mu$m$^{-1}$, $k_y$ between
 $\mp 0.6~\mu$m$^{-1}$. With increasing time the signal intensity
increases about two times and reaches its maximum at $t\sim$ 0.6
ns without any marked narrowing in its $k$-distribution specific
for the stimulated polariton scattering under cw excitation.

\begin{figure}
\begin{center}
\includegraphics[width=0.5\linewidth]{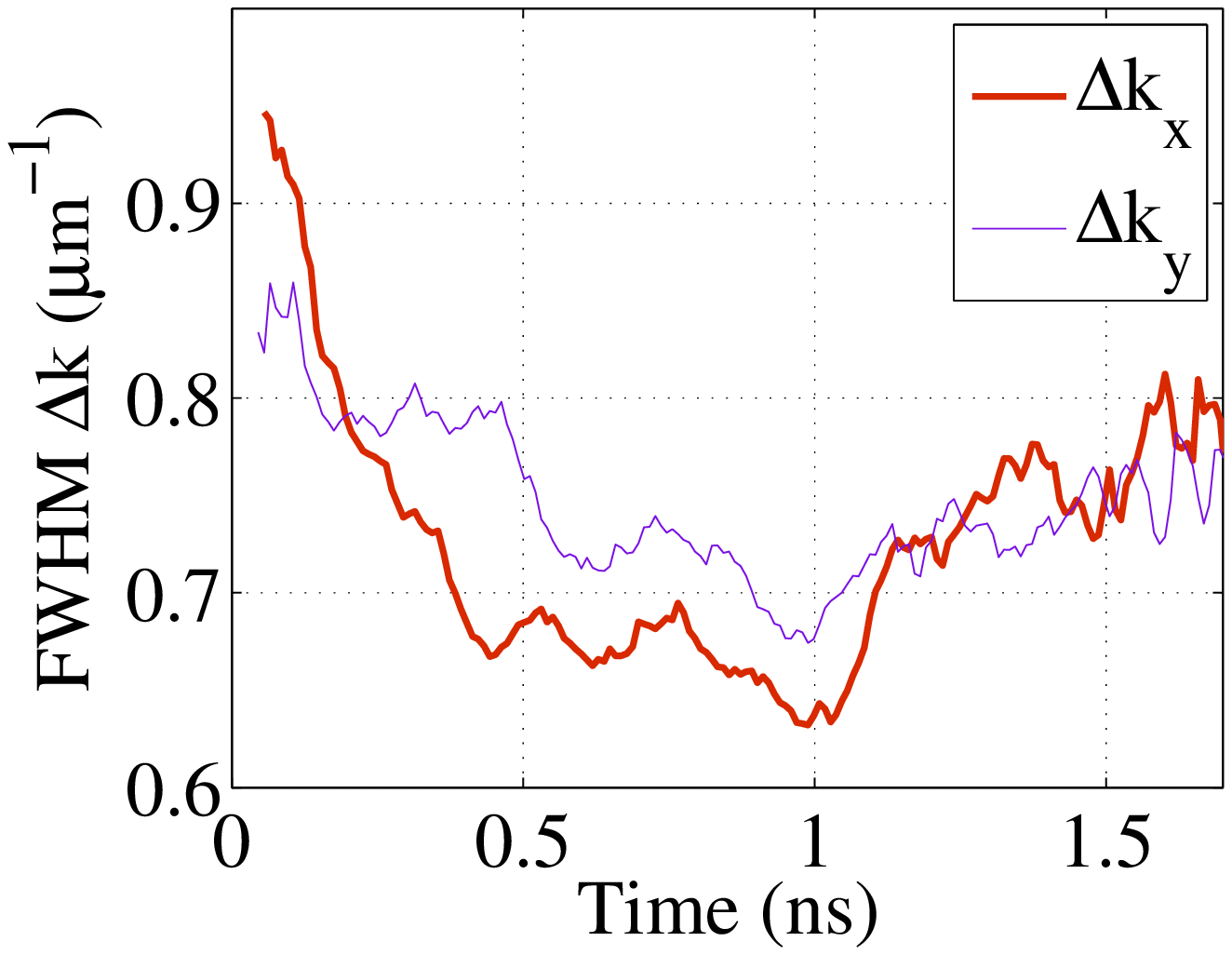}
\caption{ \label{fig4}
(Color online)  Recorded kinetics of a FWHM of $k$-distribution of LP emission
$\Delta k_x$ ($k_y=0$) and $\Delta k_y$ ($k_x=0$) at $P=16.2$~kW/cm$^2$.
}
\end{center}
\end{figure}

The marked narrowing of the signal in the $k$-space during the
excitation pulse duration of $\sim 0.8$~ns appears at higher $P$.
That is illustrated in Figs.~\ref{fig3}b and d displaying the LP
scattering dynamics at $P=10.5$~kW/cm$^2 \sim 3P_\mathrm{thr}$. The
bistable transition of the driven mode at this $P$ occurs earlier,
at $t\sim 0.1$~ns, i.e. nearly at the pump maximum and results in a
highly enhanced LP-LP scattering. The development of the
stimulated signal near the band bottom in this case is followed by
a monotonous narrowing of the $k-$distribution of the LP emission
both in $k_x$ and $k_y$ directions. Figure~\ref{fig4} displaying
the dependences of a FWHM of $k$-distribution shows that the
dynamical self-organization of the parametric scattering takes a
long time: the narrowing of the angle distribution takes place in
the whole time range of the strong scattering signal up to $t\sim
1$~ns. The $k$-distribution FWHM decreases in this time interval
from $\sim$ 1 to 0.7 $\mu$m$^{-1}$, which is still markedly larger
than the FWHM in the case of cw excitation ($\lesssim 0.3$
$\mu$m$^{-1}$). These experimental results clearly prove the DSO
OPO model.

To conclude,  the strong hysteresis effects in the kinetics of the
pumped and scattered polariton populations have been observed for
the first time in a planar semiconductor MC under a
nano-second-long pulsed resonant  excitation slightly above the LP
branch. The hysteresis effects are explained in the model of a
hard regime of the onset of parametric scattering, when the
bistability of the nonlinear pumped LP mode is accompanied by the
explosion-like growth of the scattered LP population and
subsequent dynamical self-organization process in the open
polariton system resulting in dynamically self-organized OPO.

We thank M.~S.~Skolnick for rendered samples. This work was
supported by the Russian Foundation for Basic Research, the
Russian Academy of Sciences and the ANR Chair of Excellence
Program.

%\bibliography{fwm}% Produces the bibliography via BibTeX.

\begin{thebibliography}{21}
\expandafter\ifx\csname natexlab\endcsname\relax\def\natexlab#1{#1}\fi
\expandafter\ifx\csname bibnamefont\endcsname\relax
  \def\bibnamefont#1{#1}\fi
\expandafter\ifx\csname bibfnamefont\endcsname\relax
  \def\bibfnamefont#1{#1}\fi
\expandafter\ifx\csname citenamefont\endcsname\relax
  \def\citenamefont#1{#1}\fi
\expandafter\ifx\csname url\endcsname\relax
  \def\url#1{\texttt{#1}}\fi
\expandafter\ifx\csname urlprefix\endcsname\relax\def\urlprefix{URL }\fi
\providecommand{\bibinfo}[2]{#2}
\providecommand{\eprint}[2][]{\url{#2}}

\bibitem[{\citenamefont{Stevenson et~al.}(2000)\citenamefont{Stevenson,
  Astratov, Skolnick, Whittaker, Emam-Ismail, Tartakovskii, Savvidis, Baumberg,
  and Roberts}}]{Stevenson2000}
\bibinfo{author}{\bibfnamefont{R.~M.} \bibnamefont{Stevenson}},
  \bibinfo{author}{\bibfnamefont{V.~N.} \bibnamefont{Astratov}},
  \bibinfo{author}{\bibfnamefont{M.~S.} \bibnamefont{Skolnick}},
  \bibinfo{author}{\bibfnamefont{D.~M.} \bibnamefont{Whittaker}},
  \bibinfo{author}{\bibfnamefont{M.}~\bibnamefont{Emam-Ismail}},
  \bibinfo{author}{\bibfnamefont{A.~I.} \bibnamefont{Tartakovskii}},
  \bibinfo{author}{\bibfnamefont{P.~G.} \bibnamefont{Savvidis}},
  \bibinfo{author}{\bibfnamefont{J.~J.} \bibnamefont{Baumberg}},
  \bibnamefont{and} \bibinfo{author}{\bibfnamefont{J.~S.}
  \bibnamefont{Roberts}}, \bibinfo{journal}{Phys. Rev. Lett.}
  \textbf{\bibinfo{volume}{85}}, \bibinfo{pages}{3680} (\bibinfo{year}{2000}).

\bibitem[{\citenamefont{Tartakovskii et~al.}(2000)\citenamefont{Tartakovskii,
  Krizhanovskii, and Kulakovskii}}]{Tartakovskii2000}
\bibinfo{author}{\bibfnamefont{A.~I.} \bibnamefont{Tartakovskii}},
  \bibinfo{author}{\bibfnamefont{D.~N.} \bibnamefont{Krizhanovskii}},
  \bibnamefont{and} \bibinfo{author}{\bibfnamefont{V.~D.}
  \bibnamefont{Kulakovskii}}, \bibinfo{journal}{Phys. Rev. B}
  \textbf{\bibinfo{volume}{62}}, \bibinfo{pages}{R13298}
  (\bibinfo{year}{2000}).

\bibitem[{\citenamefont{Baumberg et~al.}(2000)\citenamefont{Baumberg, Savvidis,
  Stevenson, Tartakovskii, Skolnick, Whittaker, and Roberts}}]{Baumberg2000}
\bibinfo{author}{\bibfnamefont{J.~J.} \bibnamefont{Baumberg}},
  \bibinfo{author}{\bibfnamefont{P.~G.} \bibnamefont{Savvidis}},
  \bibinfo{author}{\bibfnamefont{R.~M.} \bibnamefont{Stevenson}},
  \bibinfo{author}{\bibfnamefont{A.~I.} \bibnamefont{Tartakovskii}},
  \bibinfo{author}{\bibfnamefont{M.~S.} \bibnamefont{Skolnick}},
  \bibinfo{author}{\bibfnamefont{D.~M.} \bibnamefont{Whittaker}},
  \bibnamefont{and} \bibinfo{author}{\bibfnamefont{J.~S.}
  \bibnamefont{Roberts}}, \bibinfo{journal}{Phys. Rev. B}
  \textbf{\bibinfo{volume}{62}}, \bibinfo{pages}{R16247}
  (\bibinfo{year}{2000}).

\bibitem[{\citenamefont{Ciuti et~al.}(2001)\citenamefont{Ciuti, Schwendimann,
  and Quattropani}}]{Ciuti2001}
\bibinfo{author}{\bibfnamefont{C.}~\bibnamefont{Ciuti}},
  \bibinfo{author}{\bibfnamefont{P.}~\bibnamefont{Schwendimann}},
  \bibnamefont{and}
  \bibinfo{author}{\bibfnamefont{A.}~\bibnamefont{Quattropani}},
  \bibinfo{journal}{Phys. Rev. B} \textbf{\bibinfo{volume}{63}},
  \bibinfo{pages}{041303} (\bibinfo{year}{2001}).

\bibitem[{\citenamefont{Whittaker}(2001)}]{Whittaker2001}
\bibinfo{author}{\bibfnamefont{D.~M.} \bibnamefont{Whittaker}},
  \bibinfo{journal}{Phys. Rev. B} \textbf{\bibinfo{volume}{63}},
  \bibinfo{pages}{193305} (\bibinfo{year}{2001}).

\bibitem[{\citenamefont{Savvidis et~al.}(2001)\citenamefont{Savvidis, Ciuti,
  Baumberg, Whittaker, Skolnick, and Roberts}}]{Savvidis2001}
\bibinfo{author}{\bibfnamefont{P.~G.} \bibnamefont{Savvidis}},
  \bibinfo{author}{\bibfnamefont{C.}~\bibnamefont{Ciuti}},
  \bibinfo{author}{\bibfnamefont{J.~J.} \bibnamefont{Baumberg}},
  \bibinfo{author}{\bibfnamefont{D.~M.} \bibnamefont{Whittaker}},
  \bibinfo{author}{\bibfnamefont{M.~S.} \bibnamefont{Skolnick}},
  \bibnamefont{and} \bibinfo{author}{\bibfnamefont{J.~S.}
  \bibnamefont{Roberts}}, \bibinfo{journal}{Phys. Rev. B}
  \textbf{\bibinfo{volume}{64}}, \bibinfo{pages}{075311}
  (\bibinfo{year}{2001}).

\bibitem[{\citenamefont{Kulakovskii et~al.}(2001)\citenamefont{Kulakovskii,
  Tartakovskii, Krizhanovskii, Gippius, Skolnick, and
  Roberts}}]{Kulakovskii2001}
\bibinfo{author}{\bibfnamefont{V.~D.} \bibnamefont{Kulakovskii}},
  \bibinfo{author}{\bibfnamefont{A.~I.} \bibnamefont{Tartakovskii}},
  \bibinfo{author}{\bibfnamefont{D.~N.} \bibnamefont{Krizhanovskii}},
  \bibinfo{author}{\bibfnamefont{N.~A.} \bibnamefont{Gippius}},
  \bibinfo{author}{\bibfnamefont{M.~S.} \bibnamefont{Skolnick}},
  \bibnamefont{and} \bibinfo{author}{\bibfnamefont{J.~S.}
  \bibnamefont{Roberts}}, \bibinfo{journal}{Nanotechnology}
  \textbf{\bibinfo{volume}{12}}, \bibinfo{pages}{475} (\bibinfo{year}{2001}).

\bibitem[{\citenamefont{Butte et~al.}(2003)\citenamefont{Butte, Skolnick,
  Whittaker, Bajoni, and Roberts}}]{Butte2003}
\bibinfo{author}{\bibfnamefont{R.}~\bibnamefont{Butte}},
  \bibinfo{author}{\bibfnamefont{M.~S.} \bibnamefont{Skolnick}},
  \bibinfo{author}{\bibfnamefont{D.~M.} \bibnamefont{Whittaker}},
  \bibinfo{author}{\bibfnamefont{D.}~\bibnamefont{Bajoni}}, \bibnamefont{and}
  \bibinfo{author}{\bibfnamefont{J.~S.} \bibnamefont{Roberts}},
  \bibinfo{journal}{Phys. Rev. B} \textbf{\bibinfo{volume}{68}},
  \bibinfo{pages}{115325} (\bibinfo{year}{2003}).

\bibitem[{\citenamefont{Gippius et~al.}(2004)\citenamefont{Gippius, Tikhodeev,
  Kulakovskii, Krizhanovskii, and Tartakovskii}}]{Gippius2004}
\bibinfo{author}{\bibfnamefont{N.~A.} \bibnamefont{Gippius}},
  \bibinfo{author}{\bibfnamefont{S.~G.} \bibnamefont{Tikhodeev}},
  \bibinfo{author}{\bibfnamefont{V.~D.} \bibnamefont{Kulakovskii}},
  \bibinfo{author}{\bibfnamefont{D.~N.} \bibnamefont{Krizhanovskii}},
  \bibnamefont{and} \bibinfo{author}{\bibfnamefont{A.~I.}
  \bibnamefont{Tartakovskii}}, \bibinfo{journal}{Europhys. Lett.}
  \textbf{\bibinfo{volume}{67}}, \bibinfo{pages}{997} (\bibinfo{year}{2004}).
 % \urlprefix\url{url={http://stacks.iop.org/0295-5075/67/997},}.

\bibitem[{\citenamefont{Kulakovskii et~al.}(2003)\citenamefont{Kulakovskii,
  Krizhanovskii, Tartakovskii, Gippius, and Tikhodeev}}]{Kulakovskii2003}
\bibinfo{author}{\bibfnamefont{V.~D.} \bibnamefont{Kulakovskii}},
  \bibinfo{author}{\bibfnamefont{D.~N.} \bibnamefont{Krizhanovskii}},
  \bibinfo{author}{\bibfnamefont{A.~I.} \bibnamefont{Tartakovskii}},
  \bibinfo{author}{\bibfnamefont{N.~A.} \bibnamefont{Gippius}},
  \bibnamefont{and} \bibinfo{author}{\bibfnamefont{S.~G.}
  \bibnamefont{Tikhodeev}}, \bibinfo{journal}{Physics -- Uspekhi}
  \textbf{\bibinfo{volume}{46}}, \bibinfo{pages}{967} (\bibinfo{year}{2003}),
  \bibinfo{note}{[Uspekhi Fiz. Nauk {\bf 173}, 995 (2003)]}.

\bibitem[{\citenamefont{Gippius and Tikhodeev}(2004)}]{Gippius2004a}
\bibinfo{author}{\bibfnamefont{N.~A.} \bibnamefont{Gippius}} \bibnamefont{and}
  \bibinfo{author}{\bibfnamefont{S.~G.} \bibnamefont{Tikhodeev}},
  \bibinfo{journal}{J. Phys.: Condens. Matter} \textbf{\bibinfo{volume}{16}},
  \bibinfo{pages}{S3653} (\bibinfo{year}{2004}).
%  \urlprefix\url{http://stacks.iop.org/0953-8984/16/S3653}.

\bibitem[{\citenamefont{Gippius et~al.}(2005)\citenamefont{Gippius, Tikhodeev,
  Keldysh, and Kulakovskii}}]{Gippius2005a}
\bibinfo{author}{\bibfnamefont{N.~A.} \bibnamefont{Gippius}},
  \bibinfo{author}{\bibfnamefont{S.~G.} \bibnamefont{Tikhodeev}},
  \bibinfo{author}{\bibfnamefont{L.~V.} \bibnamefont{Keldysh}},
  \bibnamefont{and} \bibinfo{author}{\bibfnamefont{V.~D.}
  \bibnamefont{Kulakovskii}}, \bibinfo{journal}{Physics -- Uspekhi}
  \textbf{\bibinfo{volume}{48}}, \bibinfo{pages}{306} (\bibinfo{year}{2005})
  \bibinfo{note}{[Uspekhi Fiz. Nauk {\bf 175}, 327 (2005)]}.

\bibitem[{\citenamefont{Baas et~al.}(2004{\natexlab{a}})\citenamefont{Baas,
  Karr, Eleuch, and Giacobino}}]{Baas2004}
\bibinfo{author}{\bibfnamefont{A.}~\bibnamefont{Baas}},
  \bibinfo{author}{\bibfnamefont{J.~P.} \bibnamefont{Karr}},
  \bibinfo{author}{\bibfnamefont{H.}~\bibnamefont{Eleuch}}, \bibnamefont{and}
  \bibinfo{author}{\bibfnamefont{E.}~\bibnamefont{Giacobino}},
  \bibinfo{journal}{Phys. Rev. A} \textbf{\bibinfo{volume}{69}},
  \bibinfo{pages}{023809} (\bibinfo{year}{2004}{\natexlab{a}}).

\bibitem[{\citenamefont{Baas et~al.}(2004{\natexlab{b}})\citenamefont{Baas,
  Karr, Romanelli, Bramati, and Giacobino}}]{Baas2004a}
\bibinfo{author}{\bibfnamefont{A.}~\bibnamefont{Baas}},
  \bibinfo{author}{\bibfnamefont{J.-P.} \bibnamefont{Karr}},
  \bibinfo{author}{\bibfnamefont{M.}~\bibnamefont{Romanelli}},
  \bibinfo{author}{\bibfnamefont{A.}~\bibnamefont{Bramati}}, \bibnamefont{and}
  \bibinfo{author}{\bibfnamefont{E.}~\bibnamefont{Giacobino}},
  \bibinfo{journal}{Phys. Rev. B} \textbf{\bibinfo{volume}{70}},
  \bibinfo{pages}{161307(R)} (\bibinfo{year}{2004}{\natexlab{b}}).

\bibitem[{\citenamefont{Whittaker}(2005)}]{Whittaker2005}
\bibinfo{author}{\bibfnamefont{D.~M.} \bibnamefont{Whittaker}},
  \bibinfo{journal}{Phys. Rev. B}
  \textbf{\bibinfo{volume}{71}}, \bibinfo{eid}{115301}
  %(pages~\bibinfo{numpages}{7})
   (\bibinfo{year}{2005}).
 % \urlprefix\url{http://link.aps.org/abstract/PRB/v71/e115301}.

\bibitem[{\citenamefont{Wouters and Carusotto}(2007)}]{Wouters2007}
\bibinfo{author}{\bibfnamefont{M.}~\bibnamefont{Wouters}} \bibnamefont{and}
  \bibinfo{author}{\bibfnamefont{I.}~\bibnamefont{Carusotto}},
  \bibinfo{journal}{Phys. Rev. B }
  \textbf{\bibinfo{volume}{75}}, \bibinfo{eid}{075332}
  %(pages~\bibinfo{numpages}{12})
  (\bibinfo{year}{2007}).
 % \urlprefix\url{http://link.aps.org/abstract/PRB/v75/e075332}.

\bibitem[{\citenamefont{Gavrilov et~al.}(2007)\citenamefont{Gavrilov, Gippius,
  Kulakovskii, and Tikhodeev}}]{Gavrilov2007}
\bibinfo{author}{\bibfnamefont{S.~S.} \bibnamefont{Gavrilov}},
  \bibinfo{author}{\bibfnamefont{N.~A.} \bibnamefont{Gippius}},
  \bibinfo{author}{\bibfnamefont{V.~D.} \bibnamefont{Kulakovskii}},
  \bibnamefont{and} \bibinfo{author}{\bibfnamefont{S.~G.}
  \bibnamefont{Tikhodeev}}, \bibinfo{journal}{Zh. Eksp. Teor. Fiz.}
  \textbf{\bibinfo{volume}{131}}, \bibinfo{pages}{819} (\bibinfo{year}{2007}),
  \bibinfo{note}{[JETP \textbf{104}, 715 (2007)]}.

\bibitem[{\citenamefont{Langbein}(2004)}]{Langbein2004}
\bibinfo{author}{\bibfnamefont{W.}~\bibnamefont{Langbein}},
  \bibinfo{journal}{Phys. Rev. B} \textbf{\bibinfo{volume}{70}},
  \bibinfo{pages}{205301} (\bibinfo{year}{2004}).

\bibitem[{\citenamefont{Krizhanovskii et~al.}(2007)\citenamefont{Krizhanovskii,
  Gavrilov, Love, Sanvitto, Gippius, Tikhodeev, Kulakovskii, Whittaker,
  Skolnick, and Roberts}}]{Krizhanovskii2007}
\bibinfo{author}{\bibfnamefont{D.~N.} \bibnamefont{Krizhanovskii}},
  \bibinfo{author}{\bibfnamefont{S.~S.} \bibnamefont{Gavrilov}},
  \bibinfo{author}{\bibfnamefont{A.~P.~D.} \bibnamefont{Love}},
  \bibinfo{author}{\bibfnamefont{D.}~\bibnamefont{Sanvitto}},
  \bibinfo{author}{\bibfnamefont{N.~A.} \bibnamefont{Gippius}},
  \bibinfo{author}{\bibfnamefont{S.~G.} \bibnamefont{Tikhodeev}},
  \bibinfo{author}{\bibfnamefont{V.~D.} \bibnamefont{Kulakovskii}},
  \bibinfo{author}{\bibfnamefont{D.~M.} \bibnamefont{Whittaker}},
  \bibinfo{author}{\bibfnamefont{M.~S.} \bibnamefont{Skolnick}},
  \bibnamefont{and} \bibinfo{author}{\bibfnamefont{J.~S.}
  \bibnamefont{Roberts}}, \bibinfo{journal}{submitted}  (\bibinfo{year}{2007}).

\bibitem[{\citenamefont{Bloch and Marzin}(1997)}]{Bloch1997}
\bibinfo{author}{\bibfnamefont{J.}~\bibnamefont{Bloch}} \bibnamefont{and}
  \bibinfo{author}{\bibfnamefont{J.~Y.} \bibnamefont{Marzin}},
  \bibinfo{journal}{Phys. Rev. B} \textbf{\bibinfo{volume}{56}},
  \bibinfo{pages}{2103} (\bibinfo{year}{1997}).

\bibitem[{\citenamefont{Tassone et~al.}(1997)\citenamefont{Tassone,
  Piermarocchi, Savona, Quattropani, and Schwendimann}}]{Tassone1997}
\bibinfo{author}{\bibfnamefont{F.}~\bibnamefont{Tassone}},
  \bibinfo{author}{\bibfnamefont{C.}~\bibnamefont{Piermarocchi}},
  \bibinfo{author}{\bibfnamefont{V.}~\bibnamefont{Savona}},
  \bibinfo{author}{\bibfnamefont{A.}~\bibnamefont{Quattropani}},
  \bibnamefont{and}
  \bibinfo{author}{\bibfnamefont{P.}~\bibnamefont{Schwendimann}},
  \bibinfo{journal}{Phys. Rev. B} \textbf{\bibinfo{volume}{56}},
  \bibinfo{pages}{7554} (\bibinfo{year}{1997}).

\end{thebibliography}

%\end{document}

\end{document}